# THERMODYNAMIC EQUILIBRIUM IN OPEN CHEMICAL SYSTEMS.

*The R-modynamics of chemical equilibria.*

B. Zilbergleyt,
e-mail: livent1@msn.com

ABSTRACT.

The article presents new model of equilibrium in open chemical systems suggesting a linear dependence of the reaction shift from equilibrium in presence of the external thermodynamic force. Basic equation of this model contains traditional logarithmic term and a non-traditional parabolic term.. At isolated equilibrium the non-traditional term equals to zero turning the whole equation to the traditional form of constant equation. This term coincides with the excessive thermodynamic function revealing linear relationship between logarithm of the thermodynamic activity coefficient and reaction extent at open equilibrium. Discovered relationship prompts us to use in many systems a combination of the linearity coefficient and reaction shift from true equilibrium rather then activity coefficients. The coefficient of linearity can be found by thermodynamic computer simulation while the shift is an independent variable defining the open equilibrium state. Numerical data obtained by various simulation techniques proved premise of the method of chemical dynamics.

## INTRODUCTION: BACK TO CHEMICAL DYNAMICS.

Nowadays we know that chemical self-organization happens in a vaguely defined area "far-from-equilibrium"[1], while classical thermodynamics defines what is frozen at the point of true equilibrium. What occurs in between?

"True", or "internal" thermodynamic equilibrium is defined by current thermodynamic paradigm only for isolated systems. That's why applications to real systems often lead to severe misinterpretation of their status, bringing approximate rather than precise results. A few questions arise in this relation. Is it possible to expand the idea of thermodynamic equilibrium to open systems? How to describe and simulate open equilibrium in chemical systems? Is there any relationship between deviation of a chemical system from "true" equilibrium and parameters of its non-ideality?

Traditional methods use excessive thermodynamic functions to account external interaction of some system's components. It is noteworthy that the functions and related coefficients of thermodynamic activity were introduced rather for convenience [2], first playing a role of fig leave for the lack of our knowledge of what's going on in real systems.

One of the current methods in equilibrium thermodynamics of open systems, to a certain extent influenced this work, was offered by D. Korzhinsky [3]. Considering interaction of open systems within the *multisystem*, the method distinguishes between the common, or *mobile* components and specific for each subsystem *inert* components, which cannot be present in any other subsystem. The mobile components are responsible for the subsystems interaction and carry intensive thermodynamic characteristics, thus contributing the subsystem's Gibbs's potential. It is important that coefficients of thermodynamic activity of the mobile components may vary while for the inert components they do not have any physical sense [4]. The model successfully resulted in well developed theory of multisystems with extensive application output [5]. In the Korzhynsky's model openness of the system is simulated using two-level component stratification and appropriate expression for change of Gibbs' potential. Interaction with other parts of the multisystem in this model may be simulated via series of consecutive titrations of the subsystem by mobile components.

To answer the above questions more consistently, we used currently almost neglected de Donder's method. T. De Donder has introduced the thermodynamic affinity, interpreting it as a thermodynamic force and considering the reaction extent a "chemical distance" [6]. For greater convenience and universalization of the method, we have redefined the reaction extent as $d\xi_j=dn_{kj}/\eta_{kj}$, instead of $d\xi_j=dn_{kj}/\nu_{kj}$ by de Donder, or $\Delta\xi_j =\Delta n_{kj}/\eta_{kj}$ in increments. Value of $\Delta n_{kj}$ equals to amount of moles, consumed or appeared in j-reaction between its two arbitrary states, one of them usually is the initial state. The $\eta_{ij}$ value equals to a number of moles of k-component, consumed or appeared in an *isolated* j-reaction on its way from initial state to true equilibrium and may be considered *a thermodynamic equivalent of chemical transformation*. Thus redefined value of the reaction extent remains the same being calculated for any component of a simple chemical reaction; the only (and easily achievable by appropriate choice of the basis of the chemical system) condition for this is that each chemical element is involved in only one substance on each side of the reaction equation. Now, in our definition $\Delta\xi_j$ is a dimensionless chemical distance ("cd") between initial and running states of j-reaction, $0= \Delta\xi_j=1$, and thermodynamic affinity $A = - (\Delta G/\Delta\xi)_{p,T}$ turns into a classical force by definition, customary in physics and related sciences.

Chemical reaction in *isolated system* is driven only by internal force (eugenaffinity, $A_{ij}$). True thermodynamic equilibrium occurs at $A'_{ij} = 0$, and at this point $\Delta'\xi_j = 1$. Reactions in *open system* are driven by both *internal* and *external* ($A_{ej}$) forces [7] where the external force originates from chemical or, in general, thermodynamic (also due to heat exchange, pressure, etc.) interaction of the open system with its environment. Linear constitutional equations of non-equilibrium thermodynamics at zero reaction rate give us the condition of the "open" equilibrium with resultant affinity

$$A^*_{ij} + a_{ie} A^*_{ej} = 0, \tag{1}$$

where $a_{ie}$ is the Onsager coefficient [7]. The accent mark and asterisk relate values to isolated ("true") or open equilibrium correspondingly.

In this work we will use only one assumption which in fact slightly extends the hypothesis of linearity. Taking as given that there must be a relation between the reaction shift from equilibrium $\delta\xi_j =1 - \Delta\xi_j$ and external thermodynamic force causing this shift, we suppose at the first approximation that *the reaction shift in the vicinity*☆ *of true thermodynamic equilibrium is linearly related to the shifting force*

$$\delta\xi_j = \alpha_{ie} A_{ej} . \tag{2}$$

Recalling that $A_i =-(\partial G_i /\partial\xi_i )$, or $A_i =-(\Delta G_i /\Delta\xi_i )$ and substituting (2) into (1), we will have after a simple transformation and retaining in writing only $\Delta_j$ for $\Delta\xi_j$ and $\delta_j$ for $\delta\xi_j$

$$\Delta G^*_{ij} + b_{ie}\, \delta^*_j\, \Delta^*_j = 0, \tag{3}$$

where $b_{ie} = a_{ie} /\alpha_{ie}$. Corresponding constant equation is

$$\Delta G^0_{ij} + RT\ln\Pi^*_j (\eta, \Delta^*_j) + b_{ie} (1-\Delta^*_j )\Delta^*_j = 0, \tag{4}$$

where $\Pi^*_j (\eta, \Delta^*_j)$ is the activities product with mole fractions expressed using reaction extent.

So, *as soon as chemical system becomes open, given the above assumption its Gibbs' potential and the appropriate constant equation include a non-linear, non-classical term originated due to interaction of the system with its environment*.

What opens up immediately is a similarity between the non-classical term of (4) and the well known product r·x·(1-x) from the chaotic equation [8]. To get more symmetric shape of (4) we may change

☆ "*Vicinity*" in this case is certainly not less vague than "*far-from-equilibrium*". Relevant discussion will take place later on.

it defining a new value – the "non-thermodynamic", or alternative temperature of the open system

$$T_a = b_{ie} / R, \quad (5)$$

where R is universal gas constant. The value of $T_a$ is introduced in this work for convenience and symmetry; we cannot give any explanation of its physical meaning at the moment.

The logarithmic term contains well defined thermodynamic temperature $T_t$, and (4) turns to

$$\Delta G^0{}_{ij} + RT_t \ln \Pi^*{}_j + RT_a \Delta^*{}_j\,(1-\Delta^*{}_j) = 0. \quad (6)$$

Recall well known classical expression $\Delta G^0{}_{ij} = - RT\ln K_i$. Now, dividing (6) by $(-RT_t)$, presenting the activity product at open equilibrium as $\Pi^*{}_j\,(\eta_{kj}, \Delta^*{}_j) = \Pi\{[(n^0{}_{pj} + \eta_{pj}\,\Delta^*{}_j)/\Sigma]^{\nu pj} / \Pi\{[(n^0{}_{rj} - \eta_{rj}, \Delta^*{}_j)/\Sigma]^{\nu rj}$ and equilibrium constant as $K_i = \Pi`(\eta_{kj},1)$ due to $\Delta`_j = 1$, and defining reduced temperature as $\tau = T_a / T_t$ we transform equation (6) into

$$\ln [\Pi`(\eta_{kj},1)/\ \Pi(\eta_{kj}, \Delta^*{}_j)] + \tau_j\,\Delta^*{}_j\,\delta^*{}_j = 0. \quad (7)$$

Being divided by $\Delta^*{}_j$, this equation still expresses linearity between the thermodynamic force and reaction shift

$$\{\ln [\Pi`(\eta_{kj},1)/\ \Pi(\eta_{kj}, \Delta^*{}_j)]\}/\ \Delta^*{}_j = - \tau_j\,\delta^*{}_j, \quad (8)$$

while numerator of the left part is a new expression for the thermodynamic force. Containing parameters $\eta$ and $\tau$, and variable $\Delta^*{}_j$ (or $\delta^*{}_j$), equation (7) in general can be written as

$$\phi^* = \phi\,(\Delta^*, \eta, \tau). \quad (9)$$

It is easy to see that in case of isolated system $\delta^* = 0$, $\tau = 0$ as well as the thermodynamic force equals to zero, and (7) turns to the normal constant equation

$$g^* = g\,(P, T, n^*). \quad (10)$$

We distinguish between them calling (9) the Greek and (10) - the Gibbs' (Latin) equations. For better understanding of internal relations between (9) and (10) one should recall that $\eta$, while serving as a parameter of the Greek equation, is the only output from the Gibbs' equation (because $\eta = n` - n^0$, where right side contains equilibrium and initial mole amounts).

INVESTIGATION OF THE FORCE-SHIFT RELATIONSHIP.

First, consider the force expression from equation (8). Its numerator is a logarithm of a combination of molar parts products for a given stoichiometric equation. The expression under the logarithm sign is the molar parts product for ideal system divided by the same product where $\eta_{kj}$ replaced by a product $(\Delta^*{}_j\eta_{kj})$ due to the system's shift from "true" equilibrium. Table 1 represents functions $\Pi`(\eta_{kj},1)/\ \Pi(\eta_{kj}, \Delta^*{}_j)$ for some simple chemical reactions with initial amounts of reactants A and B equal to one mole. Graphs of the reaction shifts vs. thermodynamic forces are shown at Fig. 1. One can see well expressed linearity on shift-force curves. The linearity extent depends on the $\eta$ value.

Going down to real objects, consider a model system containing a double compound $A^{\bullet}R$ and an independent reactant I (for instance, sulfur) such that I reacts only with $A^{\bullet}$, while $^{\bullet}R$ restricts reaction ability of $A^{\bullet}$ and releases in the reaction as far as $A^{\bullet}$ is consumed. Symbol $A^{\bullet}$ relates to reactant A which belongs to the system (A,I) and is open to an interaction with R. Two competing processes take place in the system - decomposition of $A^{\bullet}R$, or control reaction (C): $A^{\bullet}R = A^{\bullet} + R$, and leading reaction (L): $A^{\bullet} + I = \Sigma^*{}_L$, the right side in the last case represents a sum of products. Resulting reaction in the system is $A^{\bullet}R + I = \Sigma^*{}_L + R$.

To obtain numbers for real substances, we used thermodynamic simulation (HSC Chemistry for Windows) in the model set of substances. The Is were S, C, $H_2$, and $MeO^{\bullet}Rs$ were double oxides with symbol Me standing for Co, Ni, Fe, Sr, Ca, Pb and Mn. As restricting parts $^{\bullet}R$ were used oxides of Si, Ti, Cr, and some others. Chosen double compounds had relatively high negative standard change of Gibbs' potential to provide negligible dissociation in absence of I. In chosen systems the C-reactions were $(MeO)^{\bullet}R = (MeO)^{\bullet} + {}^{\bullet}R$, and L-reactions - $(MeO)^{\bullet} + I$.

Amount of the MeO moles consumed in isolated (MeO+I) reaction between initial state and true equilibrium was taken as value of $\eta_{kL}$. Reaction extents for open L-reactions with different $^{\bullet}Rs$ have been calculated as quotients of consumed amounts of $(MeO)^{\bullet}$ (that is $\Delta n_{kj}$) by $\eta_{kL}$. As numerator for the thermodynamic force we used traditional $\Delta G_C$ (or even $\Delta G^0{}_C$ which does not make a big difference at moderate temperatures), and the force was equal to ($-\Delta G^0_C / \Delta^*_L$). Some of the results for reactions ($MeO^\bullet R+S$) are shown on Fig.2. In this group of reactions value of ($-\Delta G^0_C / \Delta^*_L$) plays role of external thermodynamic force regarding the (MeO+S) reaction.

Table 1.

Thermodynamic forces $\{\ln[\Pi`(\eta_{kj},1)/ \Pi(\eta_{kj}, \Delta^*_j)]\}/ \Delta$ for some simple chemical reactions. Initial amounts of reactants are taken equal to 1 mole and products to zero for simplicity.

| Reaction equation. | Thermodynamic force from eq. (8). |
|---|---|
| $A + B = AB$ | $[(2\eta-\eta^2)/(1-2\eta-\eta^2)] / [(2\Delta\eta-\Delta^2\eta^2)*(1-2\Delta\eta+\Delta^2\eta^2)]$ |
| $A + 2B = AB_2$ | $(1-\Delta\eta) / (\Delta-\Delta\eta)$ |
| $2A + 2B = A_2B_2$ | $[(2-3\eta)/(2-3\Delta\eta)]^3 * [(1-2\Delta\eta)/(1-2\eta)]^4 * (1/\Delta)$ |

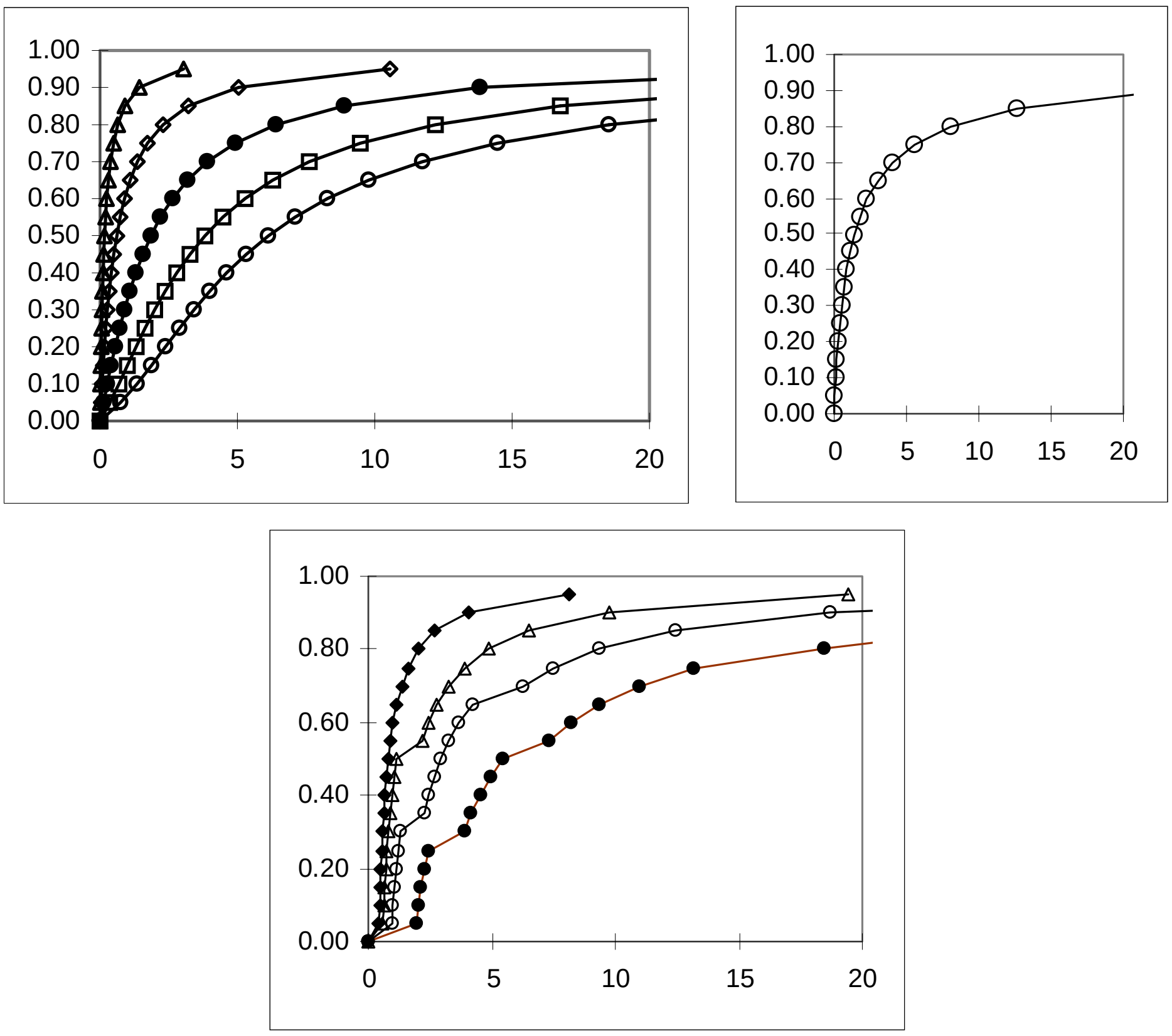

Fig. 1. Shift of some simple chemical reactions from true equilibrium (ordinate) vs. shifting force (abscissa). Reactions, left to right, values of η in brackets: A+B=AB (0.1, 0.3, .., 0.9), $A+2B=AB_2$ (0.1, 0.2, 0.3,.., 0.9), $2A+2B=A_2B_2$ (0.1, 0.2, 0.3, 0.4). One can see light delay along the x-axis for bigger η. Also, linear areas on the curves give an estimation of how far the “vicinity of equilibria” extents.

The most important is the fact that in both cases the data, showing the reality of linear relationship, have been received using exclusively current formalism of chemical equilibrium where no such kind of relationship was ever assumed at all. It is quite obvious that linear dependence took place in some cases up to essential values of deviation from equilibrium. Results shown on Fig.1 and Fig.2 prove the basics and some conclusions of the method of chemical dynamics.

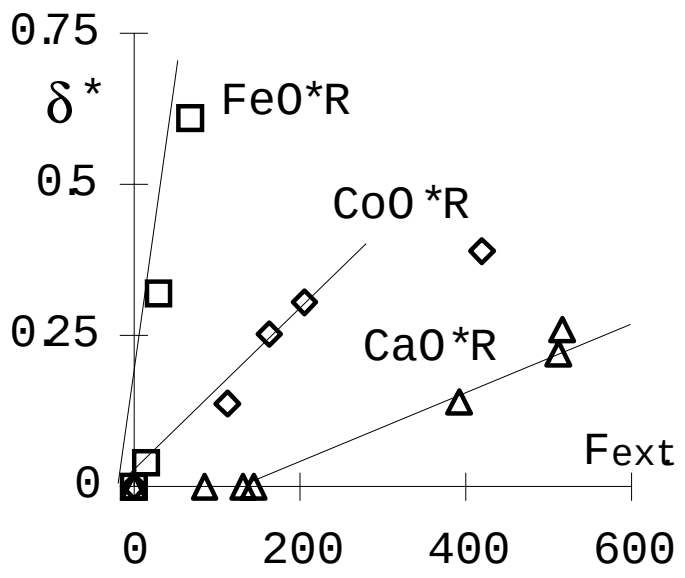

Fig.2. $\delta^*_L$ vs. force (= - $\Delta G^0_C / \Delta^*_L$ ), kJ/m cd, 298.15K, direct thermodynamic simulation. Points on the graphs correspond to various •Rs. One can see a delay along x-axis for CaO•.

FROM CHEMICAL DYNAMICS TO CHEMICAL THERMODYNAMICS:
THERMODYNAMIC ACTIVITY AND REACTION SHIFT AT OPEN EQUILIBRIUM.

It was already mentioned that classical thermodynamics has no idea of thermodynamic force. Instead, the impact of the system's interaction results in its non-ideality and usually is accounted by means of excessive thermodynamic functions and coefficients of thermodynamic activity

$$Q_j = - RT_t \ln \Pi\gamma_{kj}. \qquad (10)$$

Within current paradigm of chemical thermodynamics, constant equation for non-ideal system with $\gamma_{kj} \neq 1$ is

$$\Delta G^0_j = - RT_t \ln \Pi^*\gamma_{kj} - RT_t \cdot \ln \Pi^* x_{kj}. \qquad (11)$$

For simplicity we omitted power values, equal to stoichiometric coefficients, and $x_{kj}$ are molar fractions. The non-linear term of the Greek equation also belongs to a non-ideal system, and comparison of (6) and (11) leads to following equality in open equilibrium

$$\tau_j \,\delta^*_j = (-\ln \Pi^* \gamma_{kj})/ \Delta^*_j. \qquad (12)$$

This result is quite understandable. For instance, in case of A•R the chemical bond between A and R reduces *reaction activity* of A; the same result will be obtained for reaction (A + I) with reduced coefficient of *thermodynamic activity* of A.

Now, to avoid complexity we will be using only one common component A• in both subsystems. In this case the relationship between the L-shift and activity coefficient of A• is very simple

$$\boldsymbol{\delta^*_L = (1/ \tau_L)\, [(-\ln \gamma^*)/ \Delta^*_L]}, \qquad (14)$$

where $[(-\ln \gamma^*)/ \Delta^*_L]$ represents external thermodynamic force acting against L-reaction and divided by $RT_t$.. *This expression for the force as well as the total equation (14) are new.* This equation connects values from chemical dynamics with traditional values of classical chemical thermodynamics. Yet again, at $\delta^*_L= 0$ we have immediately $\gamma^*=1$, and vice versa, a *correlation, providing an explicit and instant transition between open and isolated systems.* In case of multiple interactions one should expect additivity of the shift increments, caused by interaction with different reaction subsystems, which follows the additively of appropriate logarithms of activity coefficients. It was also proved by simulation.

Data oon Fig. 3 were obtained using two different methods of thermodynamic simulation. I-simulation relates to an isolated (A•R+I) system with real R and A and $\gamma_{A\bullet R} =1$ in all cases. In O-simulation a combination of |A•+Y2O3+I| represented the model of open system where •R was excluded and replaced by neutral to A and I yttrium oxide to keep the same total amount of moles in the system as in I-simulation and avoid interaction between A and R. Binding of A into double compounds with R, resulting in reduced reaction ability of A, was simulated varying $\gamma_A$. I-simulation provided a relationship in corresponding rows of the $\delta^*_L$ - $\gamma^*$ values, and O-simulation - with $\delta^*_L$ - $\Delta G^0_{A\bullet R}$ correspondence. Standard change of Gibbs' potential $\Delta G^0_C$, determining strength of the A•R bond, was considered an excessive thermodynamic function to the L-reaction.

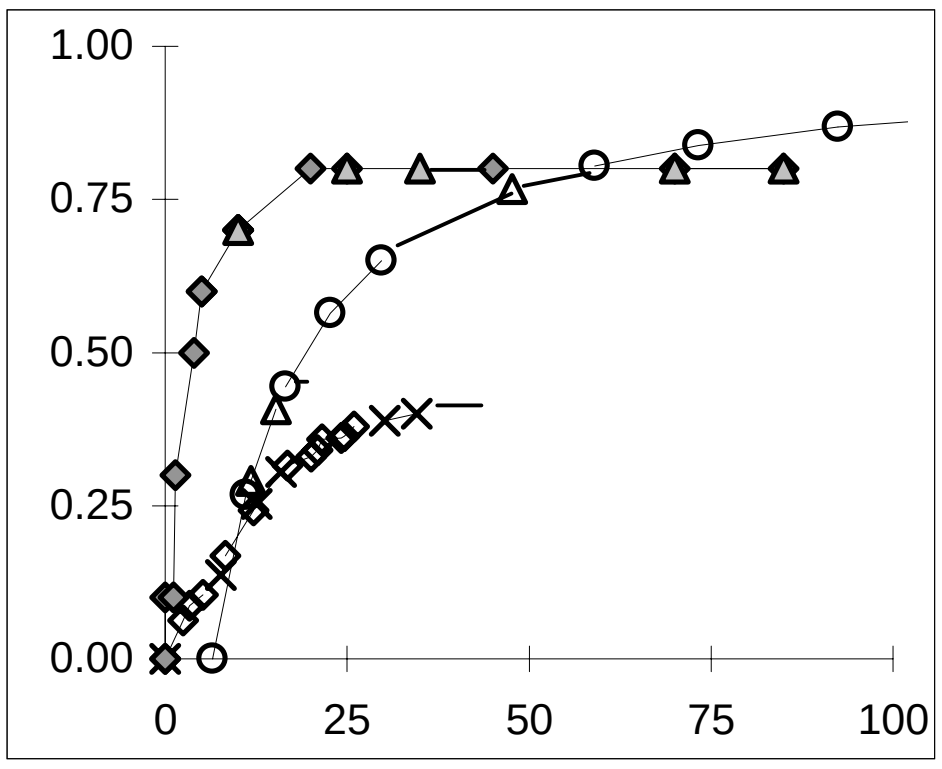

Fig. 3. $\delta$* vs. (-ln $\gamma_A/\Delta$*) (I-simulation, x) and vs. ($\Delta G^0_{A\bullet R}/\Delta$*) (O-simulation, o), (MeO$^\bullet$R+S). From left to right, PbO$^\bullet$ and CoO$^\bullet$ at 298K, SrO$^\bullet$ at 798. Curve for SrO$^\bullet$ shows light delay along the x-axis.

We have calculated some numeric values of the factor $\tau_{L_\Delta}$ from the data used for plotting Fig. 2. They are shown in Table 2.

Table 2.

Reduced temperatures, standard deviations and coefficients of determination between $\delta^*_L$ and (-ln $\gamma$*) in some A$^\bullet$R-S systems. Initial reactants ratio S/A* = 0.1.

| | CoO*R | SrO*R | PbO*R |
|---|---|---|---|
| $T_t$, K | 298.15 | 798.15 | 298.15 |
| $\tau_{L_\Delta}$ | 40.02 | 6.54 | 3.93 |
| St. deviation, % | 8.99 | 2.99 | 6.80 |
| Coeff. of determination | 0.98 | 0.99 | 0.97 |

It is worthy to mention that the range of activity coefficients usable in equilibrium calculations seems to be extendable down to unusually low powers (see Fig.1).

Strong relation between reaction shifts and activity coefficients means automatically strong relation between shifts and excessive thermodynamic functions, or external thermodynamic forces. Along with standard change of Gibbs' potential we also tried two others - the $Q_L$ which was calculated by equation (14) with $\gamma^*$, used in the O- simulation, and another, $\Delta G^*_L$ , found as a difference between $\Delta G^0_L$ and equilibrium value of $RT_t \cdot \ln\Pi^* x_{iL}$. Referring to the same $\Delta^*_L$, all three should be equal or close in values. Almost ideal match, illustrating this idea, was found in the CoO$^\bullet$R - S system and is shown on Fig.4. In other systems all three were less but still enough close. Analysis of the values, which may be used as possible excessive functions, shows that the open equilibrium may be defined using both external (like $\Delta G^0_C$) and internal (the bound affinity, see [4]) values as well as, say, a neutral, or general value like a function calculated by (14) at given activity coefficient. In principle all three may be used to calculate or evaluate $\tau_{L_\Delta}$. This allows us to reword more explicitly the problem set in the beginning of this work and explain the alternative temperature more clear. It is easy to see that equation (14) represents another form of the shift-force linearity.

Multiplying numerator and denominator of its right side by $RT_t$ and recalling that $\tau_{L_\Delta} = (T_{ch}/T_t)\Delta^*_L$ one can receive

$$\delta^*_L = [1/(RT_a)] \quad (Q_E / \Delta^*_L), \qquad (15)$$

where $Q_E$ is a general symbol for excessive thermodynamic function. *It means that the shifting force is unambiguously related to the excessive thermodynamic function, and the alternative temperature is just inverse to the coefficient of proportionality between the force and the shift it causes. The product $RT_a$ has dimension of energy while $\Delta$ and $\delta$ stay dimensionless.*

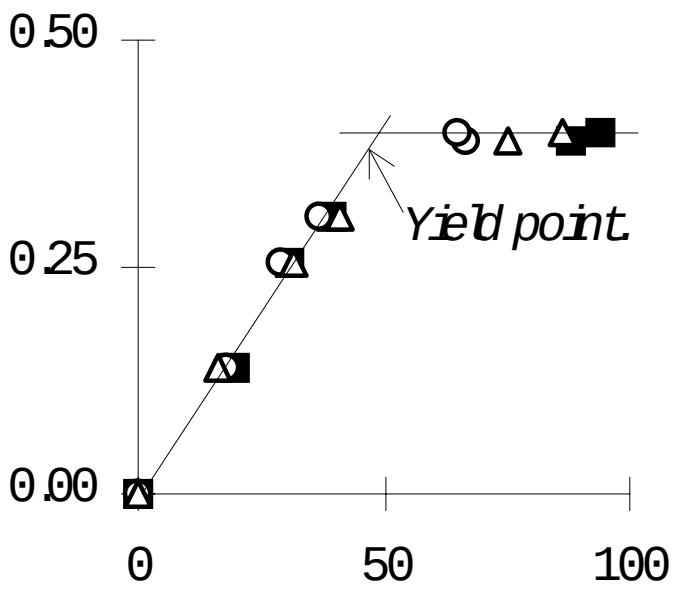

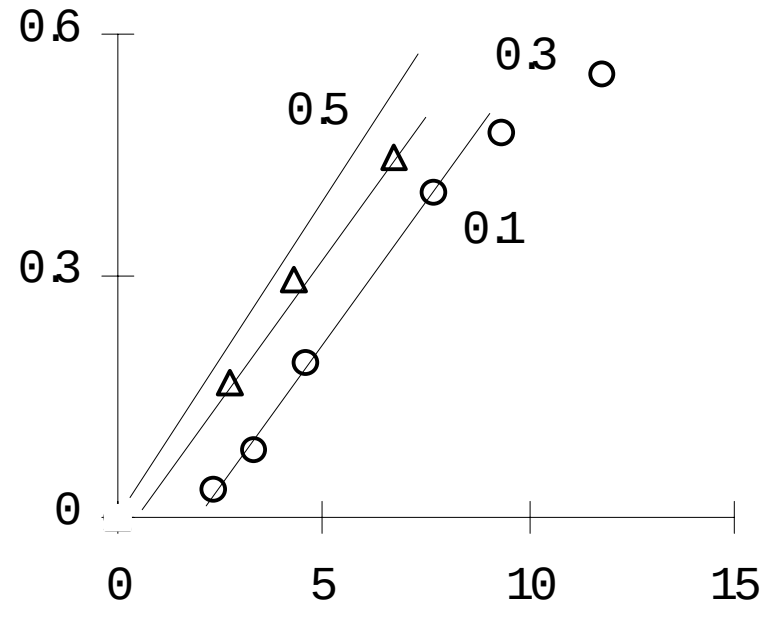

Fig. 4 (left). $\delta^*_L$ vs. shifting forces, kJ/(mole*ched). $CoO^\bullet R$-S system, 298K ($\Delta$-$Q_L/\Delta^*_L$, -$\Delta G^0_C/\Delta^*_L$, o-$\Delta G^*_L/\Delta^*_L$.).

Fig. 5 (right). $\delta^*_L$ vs. (-ln $\gamma/\Delta^*$). TDS. $CoO^\bullet R$-C system, 298K, different reactant ratio. Numbers at plots identify the initial value of the $CoO^\bullet/C$ ratio.

Because we ran thermodynamic simulation within a certain range of initial ratios between components of the reaction mixtures it was interesting to see what a difference it made. Fig. 7 shows no essential dependence of the $\tau_{L_\Delta}$ value on this ratio in the $CoO^\bullet R$ - C system.

A POSSIBILITY OF MULTIPLE STATES AT OPEN CHEMICAL EQUILIBRIUM.

Recall that equations (7) and (8) describe open equilibrium, first in terms of Gibbs' potential of the open system, second - in terms of the thermodynamic force applied to it. While the force in the "vicinity" of equilibrium is linear regarding reaction shift, the potential depends upon it parabolically thus leading to bifurcations [1] or multiple possible states of chemical system in open equilibrium.

To obtain graphs of the Greek equation, we varied $\tau$, $\eta_i$ and $\Delta^*_j$ given reaction stoichiometric equation with reasonable coefficients and initial mole amounts of reactants. One of such sets for reaction A+B = AB and $\eta_{kj}$ = 0.1 m. is shown on Fig.1.

More complicated reaction equations do not bring essential difference in the curve shapes.

Open equilibrium state in this particular model is defined essentially by standard change of Gibbs' potential $\Delta G^0_C$ of $A^\bullet R$ formation from A and R as major restricting factor. Some results of the thermodynamic simulation are plotted in coordinates $\Delta^*_L$- $\Delta G^0_C$ on Fig. 6 (marked as "Sim."). vs. calculated with equation (7) data (marked by "Calc.").

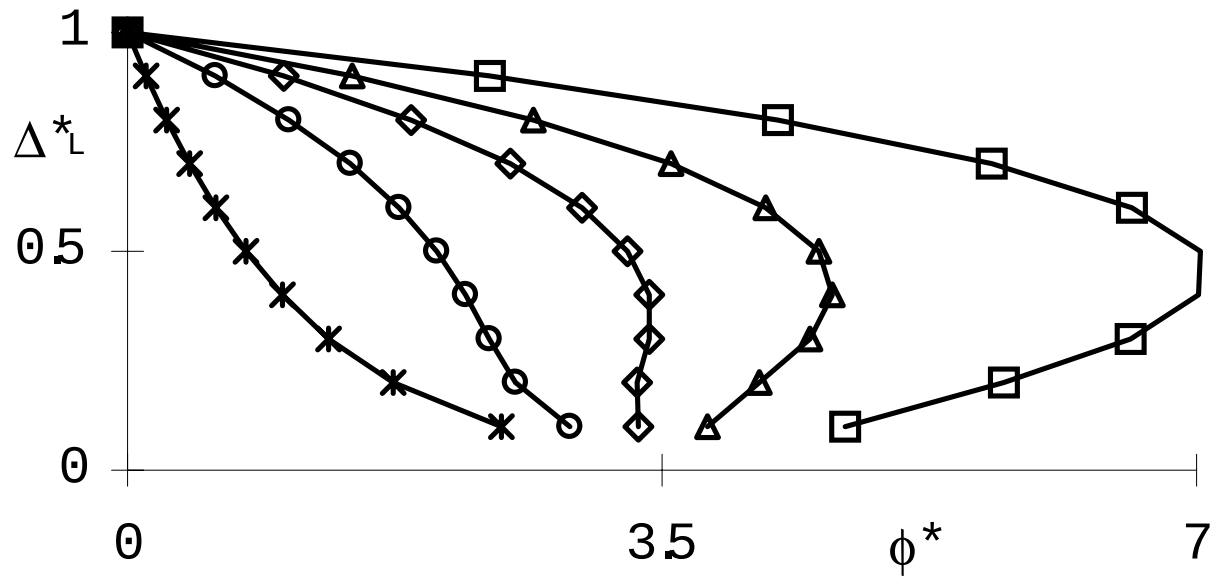

Fig.6. $\Delta^*_L$ vs. $\phi^*$. Curves for $\eta_{kj}$ = 0.1 m, $\tau$ = 0, 5, 10, 15, 25 (left to right).

Occasional scaling was applied. Points along the curves correspond to different $^\bullet$Rs. The qualitative coincidence between the curves on Fig. 7 seems to be quite satisfactory.

*Solution to the Greek equation is not unique if $\tau$ and the external thermodynamic force exceed certain values*. That leads to a very important conclusion that *the Zel`dovich's theorem* [9], *declaring the uniqueness of the state of the chemical equilibrium, is not valid beyond a certain extent of openness of chemical system and external thermodynamic forces acting against it.*

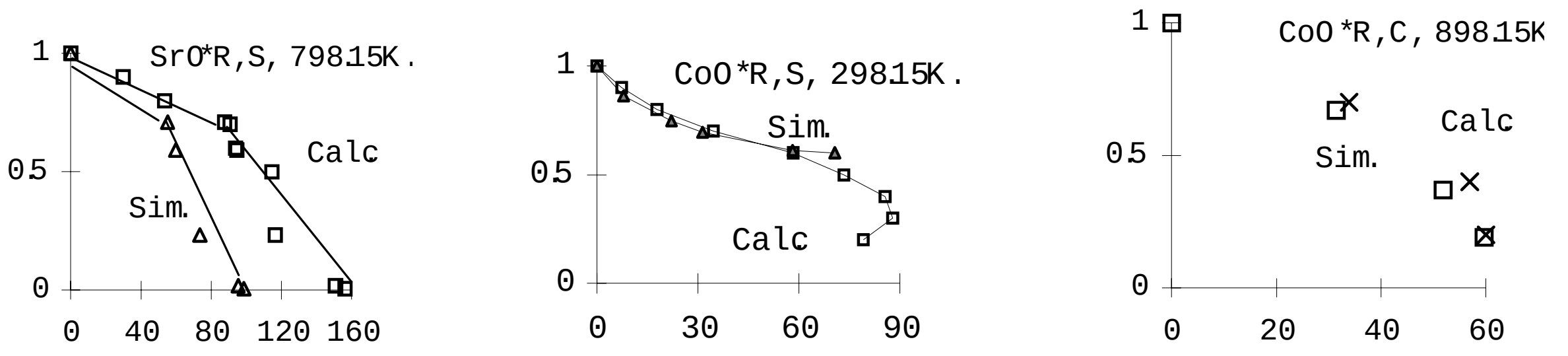

Fig.7. $\Delta^*_L$ vs. $\Delta G^0_C$ of $A^\bullet R$ formation, kJ/mol.

This part of the consequences, following from the developed method, was touched only slightly. One statement can be done for sure - the less is the value of $\eta$, that is the "weaker" is the chemical reaction the less external force is necessary to bring to multiple states at open equilibrium. It is well recognized that the bifurcations are more probable for weak reactions [1].

PRACTICAL APPLICATIONS.

We are unable to simulate and compute equilibrium composition of most complex chemical systems if we don’t know appropriate coefficients of thermodynamic activity, and their numeric values are very expensive. The method of chemical dynamics offers an easier and involving much less efforts way to run that kind of research. Indeed, equilibrium of complex chemical system per this method may be interpreted as equilibrium of subsystems’ shifts from their “true” equilibrium states, explicitly defined by the basic equation for j-subsystem. Now, having the $\eta_{kj}$ value from solution for the isolated state and $\tau_j$ as a characteristic for subsystem response to external thermodynamic perturbation (as it was above described in details), we have equation containing only $\delta^*_j$ as variable and $\tau_j$ as a parameter of the theory

$$-\ln K_j = \ln \Pi(\eta_{kj},\delta^*_j) - \tau_j \Delta^* \delta^*_j. \qquad (20)$$

Below the critical value of $\phi^*$ this equation has only one solution (see Fig. 6).

Current methods of simulation of complex equilibria use the constant equations (or equivalent expressions if minimizing Gibbs’ potential of subsystems) in the same form as if the subsystems are isolated. Their traditional joint solution is only to restrict consuming the common participants and thus achieve material balance within the system thus playing role of an accountant. Application of current methods to real systems leads to some errors in simulation results originated due to misinterpretation of their status [10]. Method developed in this work treats states of subsystems as open equilibria within a complex equilibrium, and leads to more correct numerical output.

A principal feature of application following from this method consists in usage of reaction shift (as the system’s response to external impact) multiplied by proportionality coefficient $\delta^*_j$ rather than activity coefficient $\gamma_{kj}$. Due to an easy way to obtain value of $\tau_j$ by thermodynamic simulation within minutes (not hours!), the method of chemical dynamics brings new opportunities into analysis and simulation of complex chemical systems.

DISCUSSION.

The new basic equation received in this work links equilibrium and non-equilibrium thermodynamics and may be rewritten more generally as

$$\Delta G = \Delta G^0_j + RT_t f_t(\Delta^*_j) - RT_a f_a(\Delta^*_j). \qquad (21)$$

The found relationship between reaction shift and external force resembles to a great extent the well known Hooke’s law [11] with its linearity at low elongations of a stretched material and its yield point. In our case the yield point, where the curve sharply deviates from the straight line or in some cases just changes the slope, was very distinctive on all plots. By analogy, the value of $1/t_{LD}$ may be considered *a coefficient* and the yield point - *a limit of thermodynamic proportionality* of the ***chemical Hooke’s law*** . This limit of the force-shift linearity may help to conceive the meanings of “in the vicinity of” and “far from” equilibrium areas.

We cannot tell to what extent the coefficient and the limit of thermodynamic proportionality may be considered characteristics of a *chemical elasticity* thus providing complete return of chemical reaction back to initial point when the chemical force returns to zero value, that is *without or with a sort of chemical hysteresis* in force-shift coordinates. This problem could be investigated in the future research. It must be clearly understood that despite the universality of the basic equation it would be wrong to state that *any* system may occur in the classical or non-classical areas depending on external conditions.

We call the whole method, including the original de Donder's approach, a *method of chemical dynamics*, or a *force-shift method* for explicit usage of chemical forces, originally introduced as thermodynamic affinities. The method treats true, isolated thermodynamic equilibrium of a system as a reference state for its open equilibrium when the system becomes a part of a supersystem. This reference state is memorized in $h_{kj}$. Such approach well matches interpretation of equilibrium at zero control parameters as origin of the scale of chaosity (S-theorem, [12]). Based on a very simple and quite natural assumption, *the basic equation of the present work naturally and smoothly drags non-linearity into thermodynamics of open systems thus bridging a gap between classical and non-classical thermodynamics.*

Addressing to a skeptical reader, we'd like to underline that all new results of this work have been received and proven numerically just *within the current paradigm of chemical thermodynamics.* Our non-traditional term of the basic equation already existed in chemical thermodynamics in form of excessive thermodynamic function. This work offers alternative description of its origin and its relation to an external impact on the chemical system. From this point of view, we consider results of this work neither revolutionary nor contradictory. We just tried to find out what has been lost or hidden when chemical system, the major object of chemical thermodynamics, has been idealized as an isolated entity.

And the last word – about subtitle of this article. "The R-modynamics" is "thermodynamics" (reader can easily check it) but written in a specific manner to underline that this article and the whole method of chemical dynamics are about systems with restrains, and R is traditional symbol to indicate a bind put on the system in physics, particularly in mechanics. In this connection term bound affinity was offered by the author in one of previous works [4].